\documentclass[sigconf, nonacm]{acmart}
\AtBeginDocument{%
  \providecommand\BibTeX{{%
    \normalfont B\kern-0.5em{\scshape i\kern-0.25em b}\kern-0.8em\TeX}}}

\usepackage{comment}
\usepackage{amsmath}
\usepackage{mathtools}
\usepackage{nicefrac}
\usepackage{bigints}
\usepackage{soul}
\usepackage{bigdelim}
\usepackage{arydshln}
\usepackage{picture}
\usepackage{multicol}
\usepackage{multirow}
\usepackage{romannum}

\newcommand{\smallcite}[1]{{\footnotesize\cite{#1}}}

\DeclarePairedDelimiterX{\abs}[1]{\lvert}{\rvert}{#1}

\begin{document}

\title{Estimating Discrete Total Curvature \\ with Per Triangle Normal Variation}

\author{Crane He Chen}
\email{hchen136@jhu.edu}
\affiliation{%
  \institution{Johns Hopkins University}
}

\renewcommand{\shortauthors}{Crane He Chen}


\begin{abstract}
We introduce Total Curvature Calculator, a novel approach for measuring the total curvature at every triangle of a discrete surface. This method takes advantage of the relationship between per triangle total curvature and the Dirichlet energy of the Gauss map. This new tool can be used on both triangle meshes and point clouds and has numerous applications. In this study, we demonstrate the effectiveness of our technique by using it for feature-aware mesh decimation, and show that it outperforms existing curvature-estimation methods from popular libraries such as Meshlab, Trimesh2, and Libigl. When estimating curvature on point clouds, our method outperforms popular libraries PCL and CGAL. \texttt{Libigl-}style source code is available at {\color{blue}\url{https://github.com/HeCraneChen/total-curvature-estimation.git}}. \texttt{Open3d-}style source code is available at {\color{blue}\url{https://github.com/HeCraneChen/Open3D-curvature-computation}}. 

\end{abstract}



\begin{teaserfigure}
  \begin{center}
  \includegraphics[width=\textwidth]{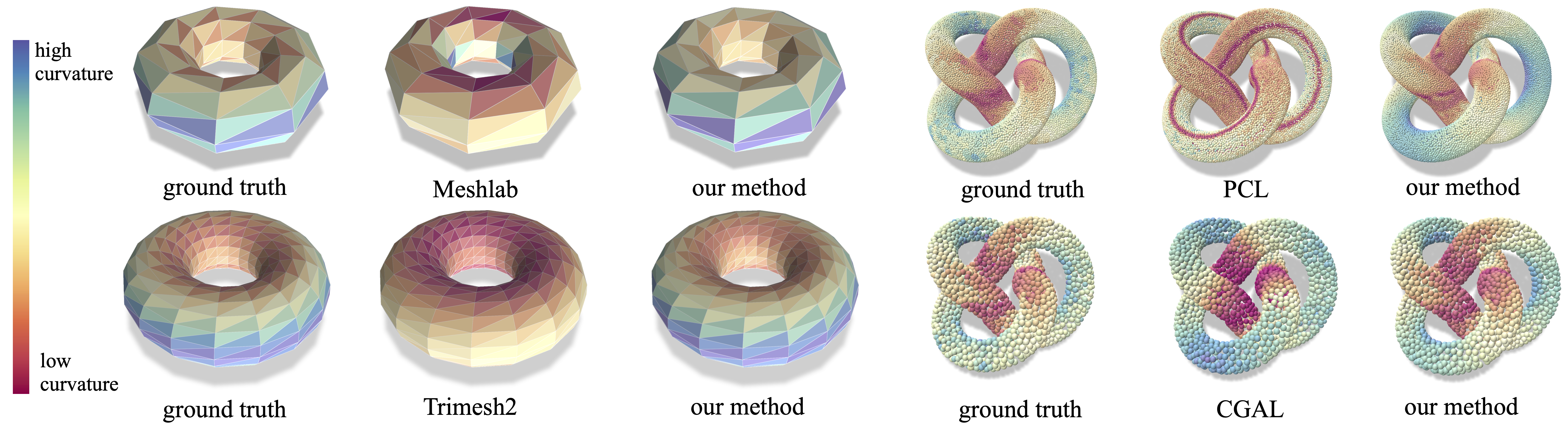}
  \caption{Comparison of discrete total curvature estimation with popular libraries. Left top: per-triangle total curvature (9x9 polyhedral torus). Left bottom: per-triangle total curvature (18x18 polyhedral torus). Right top: per point curvature averaged from total curvature (20k points). Right bottom: per point curvature averaged from total curvature (2k points)}
  \Description{teaser}
  \label{fig:teaser}
  \end{center}
\end{teaserfigure}

\maketitle

\section{Introduction}
\label{s:intro}
Curvature is an essential differential property in many geometry processing applications. In some cases, an algorithm requires the directions and values of principal curvatures. This is usually achieved by estimating a symmetric tensor approximating the shape operator. Computing the eigen-decomposition of the tensor, one obtains the principal curvature directions (the eigenvectors) and the principal curvature values (the eigenvalues). Then, curvature energies (e.g. mean, Gaussian, and total curvature) can be defined based on the estimated principal curvatures~\cite{wardetzky2007discrete}. 
\begin{figure*}[!htb]
  \includegraphics[width=\textwidth]{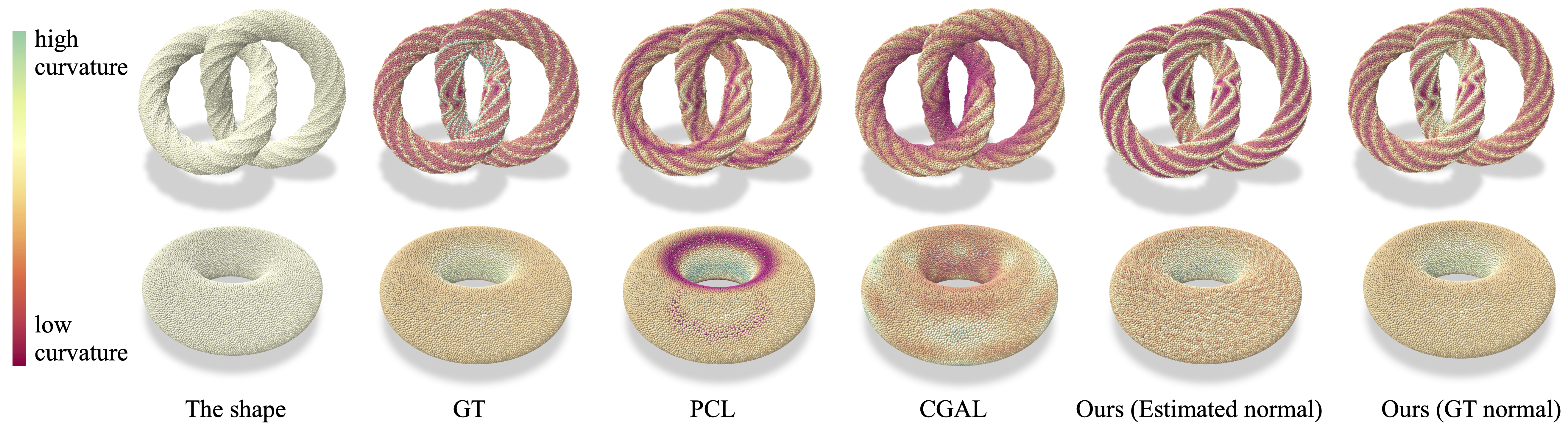}
  \caption{Comparison of curvature estimation on point clouds. (first row: knot, second row: torus)}
  \Description{pcd curvature}
  \label{fig:PCD_curvature}
\end{figure*}
We propose an alternative for directly estimating the total curvature $\kappa_1^2 + \kappa_2^2$ by integrating the variation of normal vectors, bypassing the problem of explicitly estimating the shape operator and computing its principal curvature values, $\kappa_1$ and $\kappa_2$. Specifically, our approach for total curvature estimation only requires estimation of normals and a way to compute the Dirichlet energy -- both well-studied tasks in geometry processing. Source code of \texttt{libigl-}style is freely available at {\color{blue}\url{https://github.com/HeCraneChen/total-curvature-estimation.git}}.

\section{Algorithm}
\label{s:algorithm}
Consider a triangle mesh with per vertex normals e.g. estimated by off-the-shelf algorithms. The goal is to directly estimate the total curvature over every triangle $T$

\begin{equation}
  \kappa_T = \int_T (k_1^2 + k_2^2)\,dp.
\end{equation}
Noting that the sum of the squares of the eigenvalues of a symmetric matrix equals its squared Frobenius norm and leveraging the relationship between the shape operator and the derivatives of the Gauss map, we obtain
\begin{equation}
 \kappa_T = \int_T\left\|\nabla\vec{n}\right\|_F^2\,dp
  = \sum_{i=1}^3\int_T\left\|\nabla n_i\right\|^2\,dp
 \label{eq:observation}
\end{equation}
where $\vec{n}=(n_1,n_2,n_3):T\rightarrow S^2$ is the (Gauss) map assigning a normal to every point on the triangle.\newline

The advantage of the formulation in Equation~\ref{eq:observation} is that it does not require the estimation of the shape operator. Instead, the integrals are simply the Dirichlet energies of the coordinate functions of the Gauss map -- quantities that can be computed using the cotangent Laplacian stiffness matrix.

Concretely, given a triangle $T\in\mathcal{T}$, letting $\mathbf{S}_T\in\mathbb{R}^{3\times3}$ denote the cotangent Laplacian stiffness matrix associated with triangle $T$ and setting $\mathbf{N}_T\in\mathbb{R}^{3\times3}$ to be the matrix whose column vectors are the normals at the vertices of $T$, we get:
\begin{equation*}
\kappa_T\approx \text{Trace}\left(\mathbf{N}_T\cdot\mathbf{S}_T\cdot \mathbf{N}_T^\top\right).
\end{equation*}
Note that this is an approximate estimate of the total curvature because the cotangent Laplacian assumes values are linearly interpolated from the vertices, whereas a Gauss map would require that the interpolated normal vectors be normalized to have unit-length.

Similar treatment can be applied to oriented point clouds. With normals given, all that is required is the definition of a stiffness matrix. For example, we can use the approach of Belkin~\textit{et al.}~\cite{belkin2008discrete} which defines a system matrix by constructing a local triangulation around each sample.

\section{Performance}
\label{s:performance}
For triangle meshes, we evaluate the Hausdorff distance between the decimated triangle mesh and the original triangle mesh in Table~\ref{tab:mesh}. For point clouds, we evaluate the RMSE distance between the estimated curvature and ground truth curvature quantitatively in Table~\ref{tab:pcd} and qualitatively in Figure~\ref{fig:PCD_curvature}. It can be observed that our method performs better than the methods adopted in popular libraries. For fairness of comparison, the results we show for PCL and CGAL are after carefully re-scaling using the ground truth curvature. Additionally, from Table~\ref{tab:pcd}, the quality of normal has non-negligible effects on the performance of curvature estimation. 

\begin{table}[!htb]
  \caption{Hausdorff distance between feature-aware decimated mesh and the original mesh for the bunny (top), cow (middle), and armadillo man (bottom) models.}
  \label{tab:freq}
  \begin{tabular}{ccccc}
\toprule
\multirow{2}{*}{metric} & Libigl & Meshlab & Trimesh2 & Ours \\
 & \smallcite{panozzo2010efficient} & \smallcite{taubin1995estimating} & \smallcite{rusinkiewicz2004estimating} & \\
\midrule
RMS & 0.0066 & 0.0062 & 0.0056 & \textbf{0.0054}\\
Max & 0.0542 & 0.0608 & 0.0533 & \textbf{0.0385}\\
\midrule
RMS & 0.0073 & 0.0071 & 0.0085 & \textbf{0.0069}\\
Max & 0.0731 & 0.0427 & 0.0459 & \textbf{0.0385}\\
\midrule
RMS & 0.0031 & \textbf{0.0027} & 0.0031 & \textbf{0.0027}\\
Max & 0.0370 & 0.0233 & 0.0324 & \textbf{0.0174}\\
\bottomrule
\end{tabular}

\label{tab:mesh}
\end{table}
\begin{table}[!htb]
  \caption{RMSE between ground truth curvature and estimated curvature on point clouds for the knot (top) and torus (bottom) models.}
  \label{tab:freq}
  \begin{tabular}{cccccl}
    \toprule
    \multirow{2}{*}{sampling} & PCL & CGAL & Ours  & Ours\\ && \smallcite{merigot2010voronoi}&(N est.)&(N gt)\\
    \midrule 
    uniform & 292.8847 & 342.6716 & 237.3573 & \textbf{197.5912}\\
    nonuniform & 309.8654 & 345.6605 & 295.0542 & \textbf{221.0447}\\
    sparse & 387.7908 & 438.9999 & \textbf{315.5943} & 315.7218\\
    \midrule
     uniform & 1.4364 & 1.9893 & 0.8138 & \textbf{0.0219}\\
     nonuniform & 1.5057 & 2.0118 & 1.3447 & \textbf{0.0367}\\
     sparse & 1.5792 & 2.4791 & 0.6501 & \textbf{0.0548}\\
  \bottomrule
\end{tabular}
\label{tab:pcd}
\end{table}

\section{Discussion}
\label{s:discussion}
We have introduced a simple yet effective method for total curvature estimation that is easy to integrate within existing libraries. Our results demonstrate that this method surpasses the accuracy of standard implementations that estimate the shape operator.



\appendix

\section{Appendix}
\label{a:supp_mesh}
\subsection{Additional Details--Triangle Meshes}
\begin{figure}[!htb]
  \includegraphics[width=0.5\textwidth]{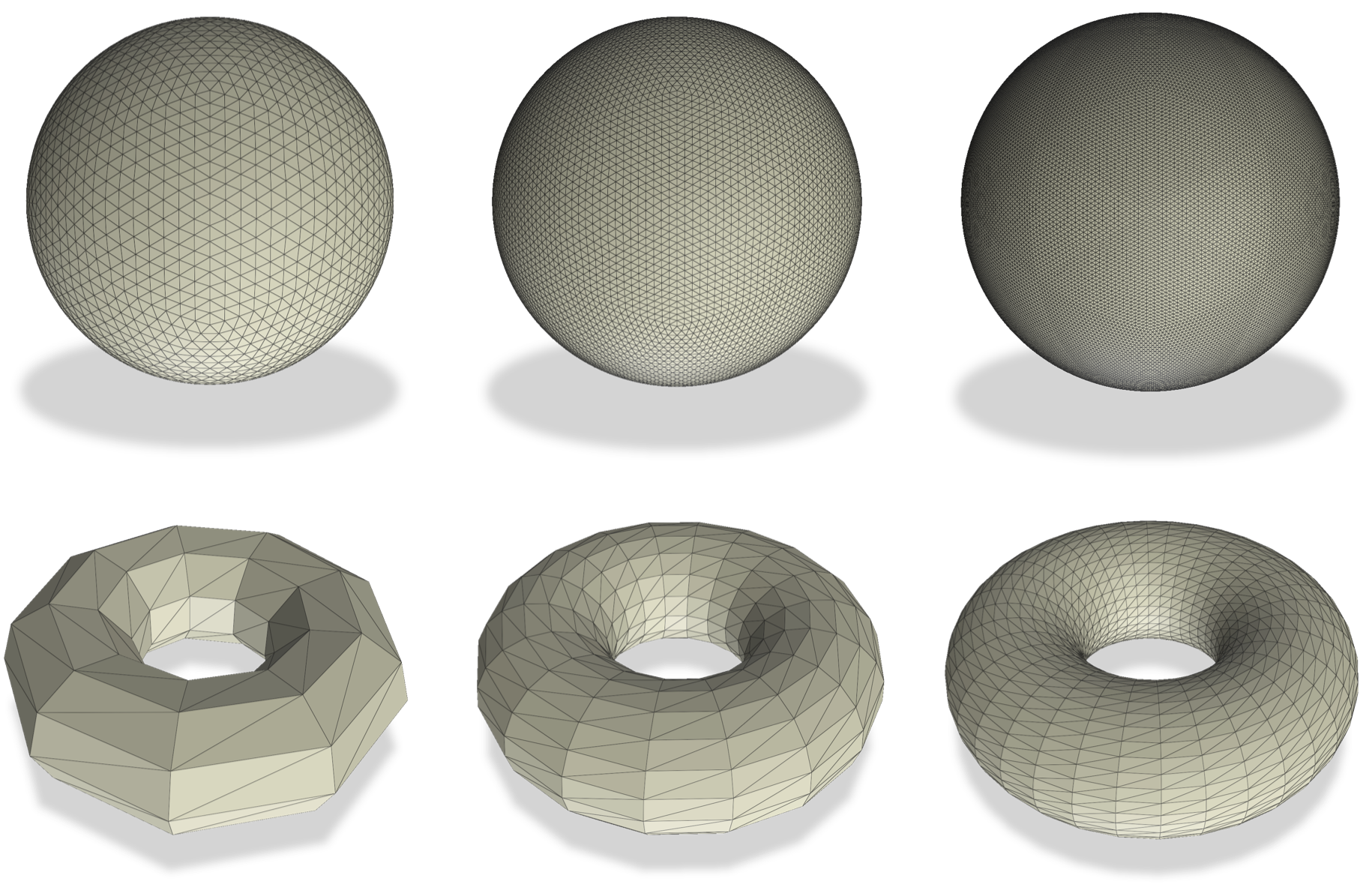}
  \caption{Meshes used for evaluation. First row: 4-subdivision, 5-subdivision, and 6-subdivision sphere from icosahedron. Second row: tori obtained by triangulating 9x9, 18x18, 36x36 grids.}
  \Description{fig:polyhedral}
  \label{fig:polyhedral}
\end{figure}

We compare results for parametric surfaces, for which an analytic expression of curvature can be obtained. Similar to Taubin~\cite{taubin1995estimating}, we evaluate total curvature estimation on the two different triangulations of a surface (icosahedron-subdivided spheres, and polyhedral tori constructed by regular grids of different resolutions). Numerical results for the meshes shown in Figure~\ref{fig:polyhedral} are presented in Table~\ref{tab:sphere_torus}. For these results, the normal vector at each point is calculated by differentiating the parameterization.

\begin{table*}[!htb]
  \caption{RMSE between ground truth and estimation of total curvature on regular triangulations of the sphere and torus at different resolutions.}
  \label{tab:freq}
  \begin{tabular}{cccccl}
    \toprule
    \multirow{2}{*}{resolution} & Libigl & Meshlab & Trimesh2 & Ours \\
 & \cite{panozzo2010efficient} & \cite{taubin1995estimating} & \cite{rusinkiewicz2004estimating} & \\
    \midrule 
    \multicolumn{5}{c}{{icosahedron-subdivided spheres}}\\
    \midrule 
    4-subdivision & 0.1104 & 0.0308 & 0.0155 & \textbf{0.0000}\\
    5-subdivision & 0.0271 & 0.0353 & 0.0155 & \textbf{0.0000}\\
    6-subdivision & 0.0067 & 0.0382 & 0.0155 & \textbf{0.0000}\\
    \midrule
    \multicolumn{5}{c}{{regularly triangulated torus}}\\
    \midrule 
     9 x 9 grid & 19.2708 & 2.5869 & 1.6643 & \textbf{0.4759}\\
     18 x 18 grid & 3.5917 & 2.6976 & 1.1838 & \textbf{0.1425}\\
     36 x 36 grid & 1.28 & 2.7072 & 1.0621 & \textbf{0.0372}\\
  \bottomrule
\end{tabular}
\label{tab:sphere_torus}
\end{table*}

\begin{table*}[!htb]
  \caption{RMSE between ground truth and estimation of total curvature on the point clouds of knots.}
  \label{tab:freq}
  \begin{tabular}{cccccl}
    \toprule
    sampling & PCL & CGAL\cite{merigot2010voronoi} & Ours (N est.) & Ours (N gt)\\
    \midrule 
    \multicolumn{5}{c}{{a torus knot}}\\
    \midrule 
    uniform & 61.7166 & 85.9137 & 25.1193 & \textbf{7.8117}\\
    nonuniform & 81.3795 & 86.0262 & 67.1373 & \textbf{8.0127}\\
    sparse & 85.7216 & 60.2592 & 28.7982 & \textbf{7.6624}\\
    \midrule
    \multicolumn{5}{c}{{another knot}}\\
    \midrule 
     uniform & 182.7609 & 218.5101 & 58.3876 & \textbf{35.484}\\
     nonuniform & 195.2599 & 243.3259 & 94.2648 & \textbf{37.1814}\\
     sparse & 208.9368 & 283.0388 & 178.3064 & \textbf{52.0716}\\
  \bottomrule
\end{tabular}
\label{tab:torus_knots}
\end{table*}
To verify and compare the efficacy of our approach on complex models, specifically those with unknown parameterizations and ground truth curvatures, we turn our attention to the mesh decimation task. This enables us to evaluate the effectiveness of the different approaches. The metric we use is the Hausdorff distance between the original mesh and the results of feature-aware decimation using curvature obtained from different estimation methods as cost function or per-vertex weight.

\begin{figure}[!htb]
  \includegraphics[width=0.5\textwidth]{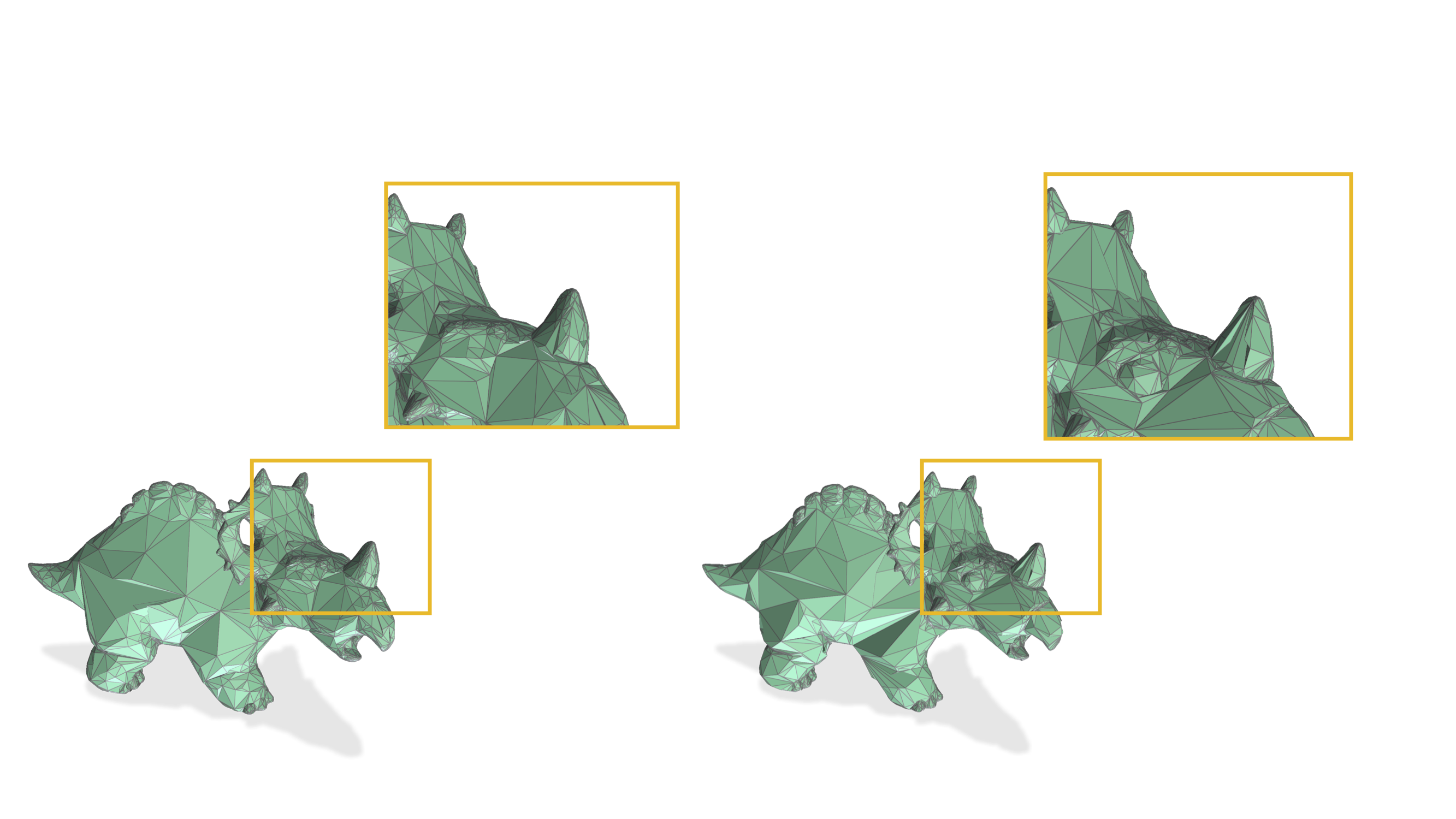}
  \caption{Comparison of curvature-weighted QSLIM decimation with total curvature calculated by different algorithms. Left: decimation using the total curvature calculated with Libigl. Right: decimation using the total curvature calculated with our method.}
  \Description{fig:decimation_comp}
  \label{fig:decimation_comp}
\end{figure}

\begin{figure}[!htb]
  \includegraphics[width=0.5\textwidth]{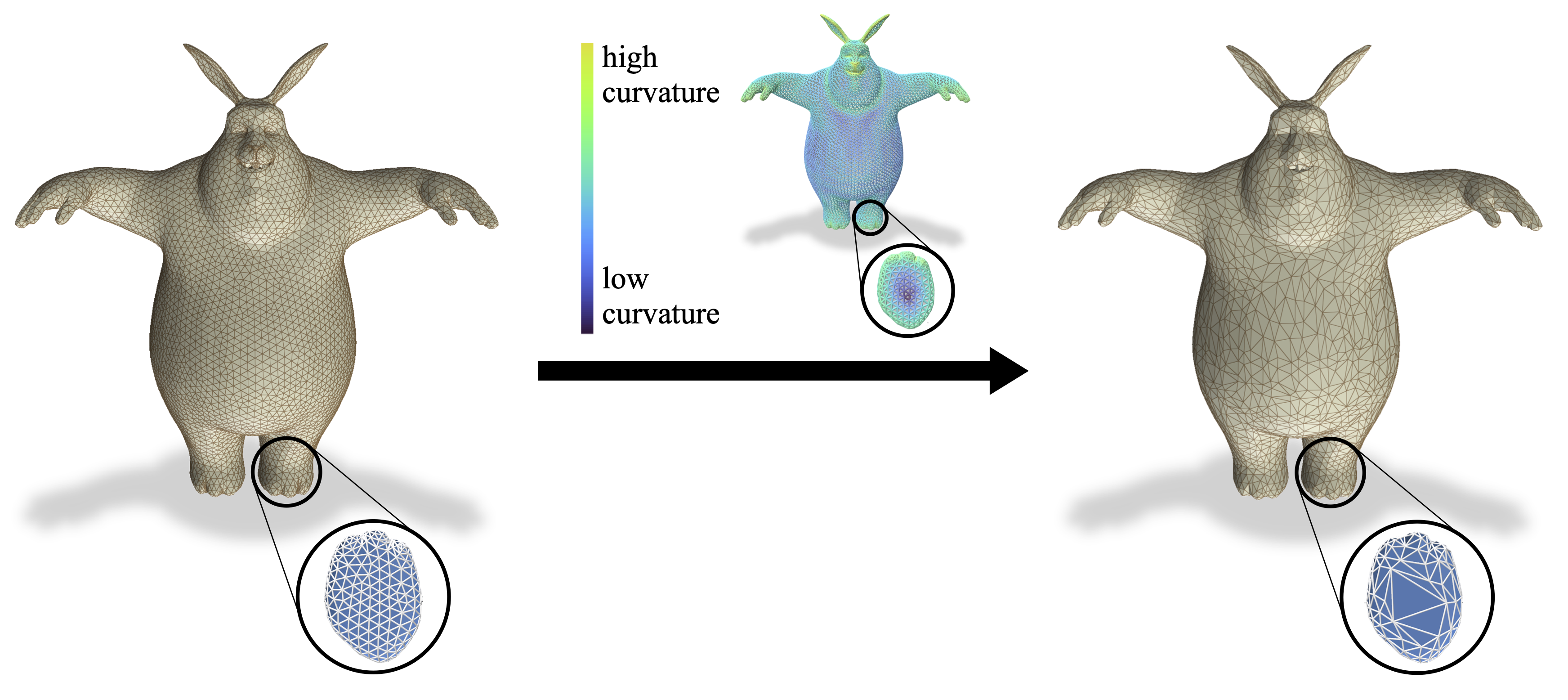}
  \caption{QSLIM-inspired feature-aware mesh decimation, where total curvature estimated by our method is used as weights. Left: before decimation. Right: after decimation.}
  \Description{fig:bunny_decimation}
  \label{fig:bunny_decimation}
\end{figure}

Here comes the implementation details. In particular, we incorporated our total curvature estimation method into two pipelines for the task, one successive method inspired by Hoppe~\cite{hoppe1996progressive} using shortest-edge-mid-point cost, and the other is a quadratic energy-based method inspired by QSLIM~\cite{garland1997surface}. In the successive methods, edge length is one of the most commonly selected cost, and its midpoint is selected as the merged vertex when edge collapsing happens. We incorporate the total curvature as a weight, which is multiplied to the edge length, to formulate a new cost function. Comparative results are shown in Figure~\ref{fig:decimation_comp}. It can be observed that highly curved regions around the arm of the mother have higher resolution, whereas conventional shortest-edge-midpoint maintains similar resolution everywhere. In the QSLIM inspired method, a total curvature weight is assigned to each vertex. The results are shown in Figure~\ref{fig:bunny_decimation}. Final results from the tables in the paper are obtained from the QSLIM inspired method. 

\begin{figure}[!htb]
  \includegraphics[width=0.5\textwidth]{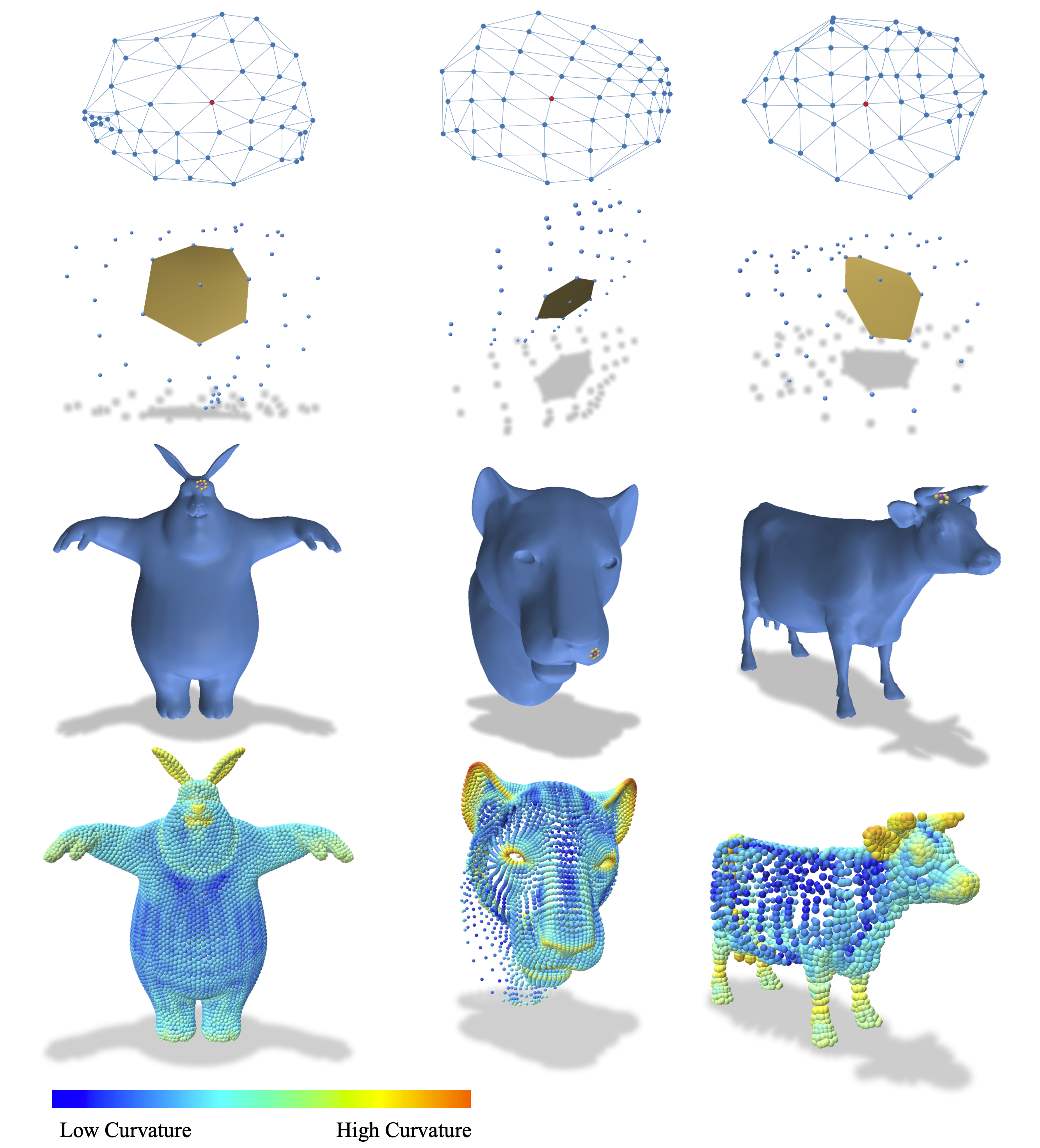}
  \caption{Curvature estimation from point clouds. Top to bottom: Delaunay Triangulation on tangent plane of the surface at the sample; local triangulation constructed with Delaunay Triangulation lifted to 3D, similar to as described in ~\cite{belkin2008discrete}; local triangles locating on the shape; Curvature estimated on point clouds. Left to right: bunny (7738 points), lion (8356 points), cow (2762 points).}
  \Description{CurvaturePCD}
  \label{fig:CurvaturePCD}
\end{figure}

\subsection{Additional Details--Point Clouds}
Our approach generalizes to point clouds, and can be implemented as follows: (1) For each point $p$, find its \textit{k}-nearest neighbors $N_k(p) = \{p_1, p_2, ..., p_k \}$, and project these points onto the tangent plane of the surface into $T_k(p) = \{p_{t1}, p_{t2}, ..., p_{tk}\}$. (2) Comute a Delaunay triangulation of $T_k(p)$, and extract the one-ring of triangles incident on $p$. (3) Calculate the curvature at $p$ by averaging the per-triangle Dirichlet energies of the 1-ring neighborhood, as in the computation of total curvature for triangle meshes. The results of this total curvature estimation are shown in Figure~\ref{fig:CurvaturePCD}.

\begin{figure}[!htb]
  \includegraphics[width=0.5\textwidth]{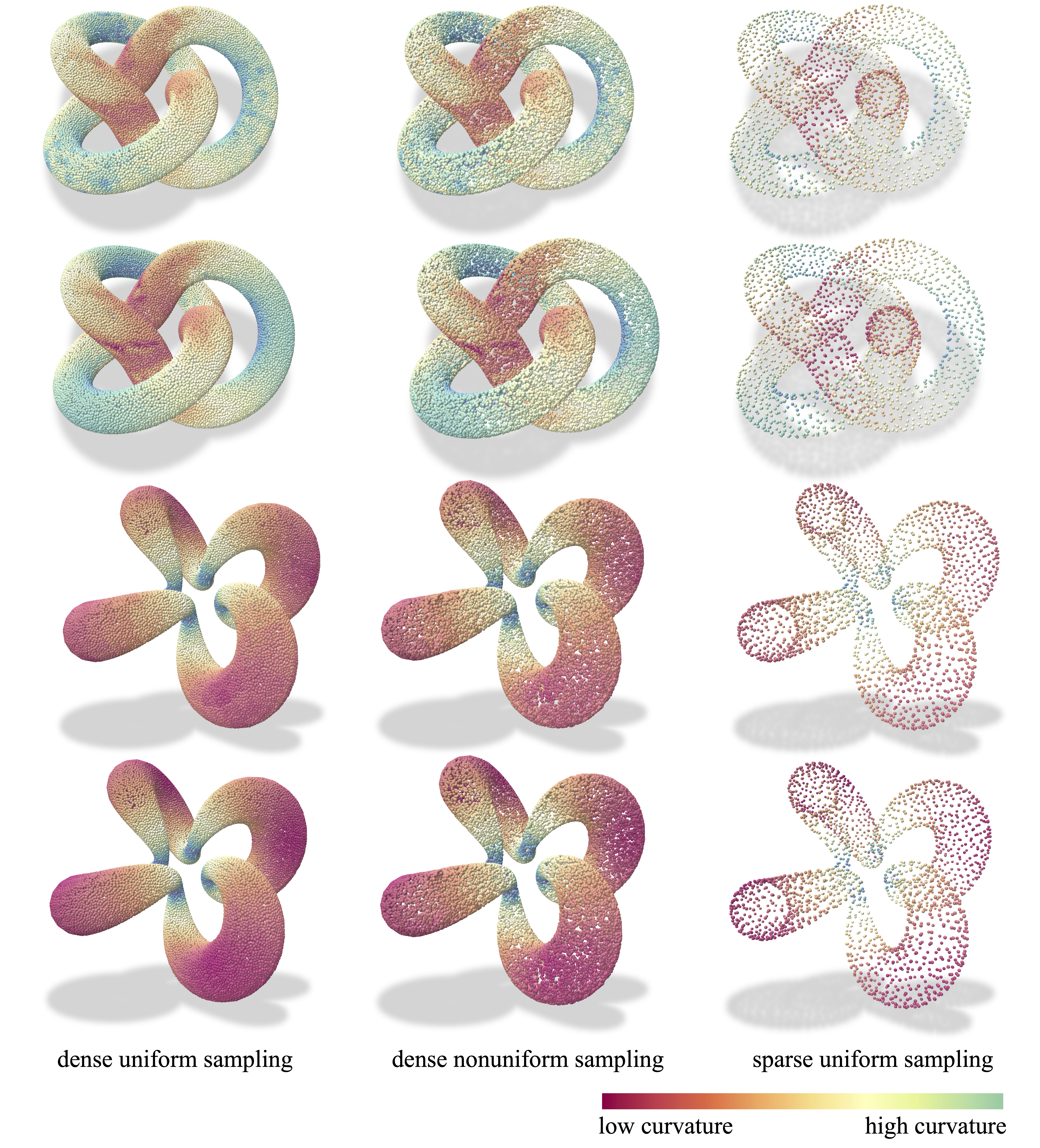}
  \caption{Comparison of curvature estimation on point cloud with respect to point sampling. First row: ground truth curvature of torus knot. Second row: curvature of torus knot estimated by our method. Third row: ground truth curvature of another knot. Fourth row: curvature of another knot estimated by our method. To isolate the problem of sampling, normals in this visualization are ground truth normals.}
  \Description{sampling_comparison}
  \label{fig:sampling_comparison}
\end{figure}
\begin{figure}[!htb]
  \includegraphics[width=0.5\textwidth]{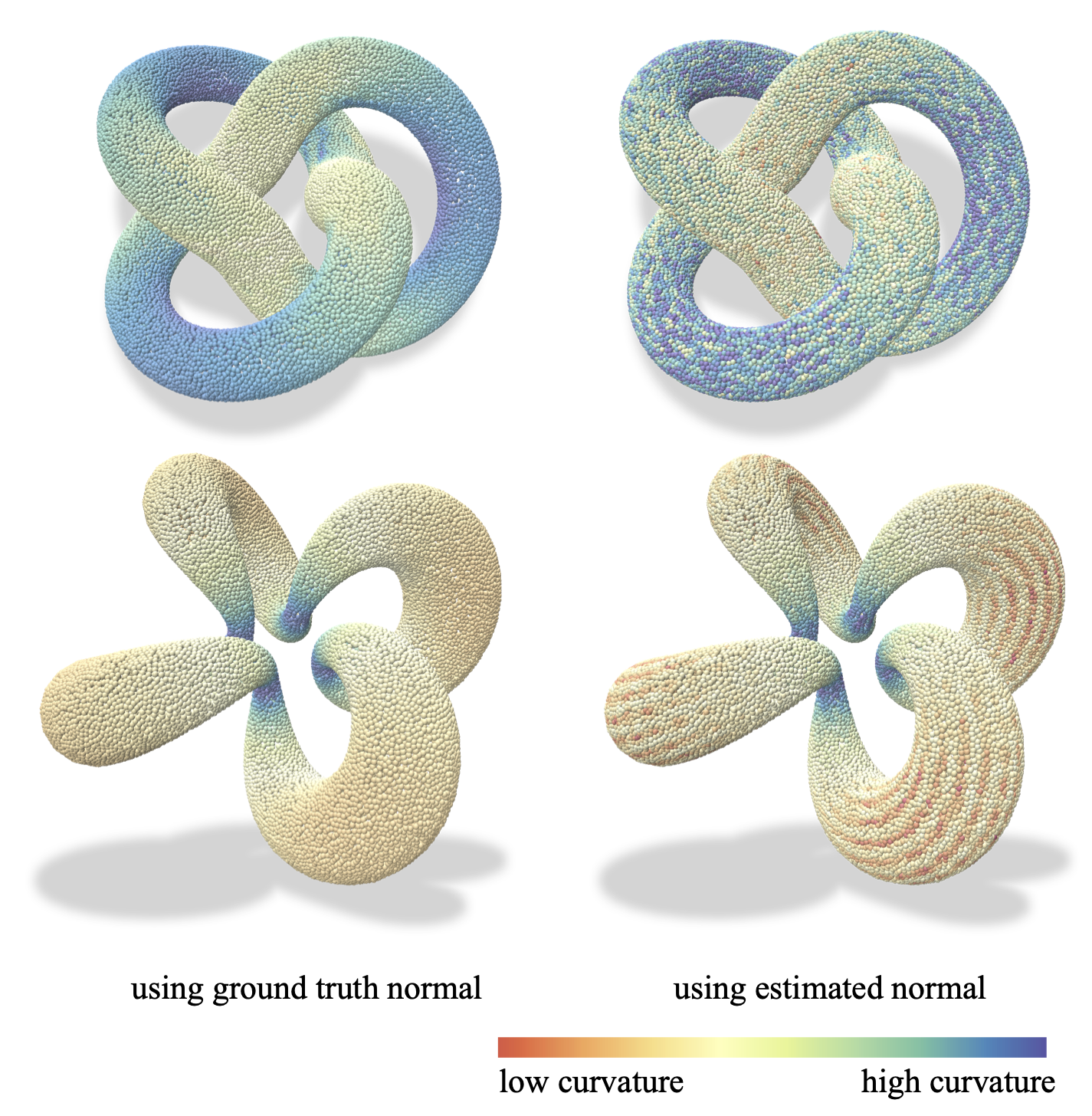}
  \caption{Comparison of curvature estimation on point cloud with respect to quality of normal. First row: torus knot. Second row: another knot.}
  \Description{normal_comparison}
  \label{fig:normal_comparison}
\end{figure}

In our results, ``uniform'' refers to a dense Poisson Disk sampling on the triangle mesh with around 20k points, ``nonuniform'' refers to first oversampling 40k points on the triangle mesh with Poisson Disk sampling, then randomly sample around 20k points form the 40k points, ``sparse'' refers to sparse Poisson Disk sampling with around 2k points.
Ground truth normals refers to the normals calculated either parametrically or estimated on the pre-known triangle mesh. Estimated normals refers to the normals estimated directly from the point clouds based on the covariance matrix of \textit{k}-nearest-neighbors. Each patch might have inconsistent sign for the normal compared to other patches. We propagate the normal orientation using a minimum spanning tree.

During the experiments, emperically, we found that compared to CGAL and PCL, our method is less sensitive to parameters. Our method takes into account the one-ring-neighborhood based on the local Delaunay triangulation. The only parameter to tune is the $k$ of \textit{k}-nearest neighbors. We select $k= 20$ for the case of dense sampling, and $k=10$ for the case of sparse sampling. Whereas both CGAL and PCL have two parameters related to radius that need to be fine-tuned in order to get good results. The parameters could differ a lot from model to model, and their selection procedure could be time-consuming. Take the torus knot as example, CGAL needs $R = 0.1, r = 0.09$ for dense sampling and $R = 0.2, r = 0.18 $ for sparse sampling. Whereas the torus in the paper works the best with $R = 0.3, r = 0.1$. Similarly, in PCL, the torus knot needs $r1 = r2 = 0.03$ for dense sampling $r1 = r2 = 0.1$ for sparse sampling, whereas the torus needs $r1 = r2 = 0.3$.

More results for point clouds are presented in Table~\ref{tab:torus_knots}. Qualitative comparisons of estimated curvature on uniform, nonuniform, and sparse point clouds with ground truth normals are shown in Figure~\ref{fig:sampling_comparison}. Qualitative comparisons for the effect of quality of normal on curvature estimation are shown in Figure~\ref{fig:normal_comparison}. It can be observed that the proposed method is robust with respect to the density and regularity of sampling, but sensitive to the quality of estimated normals.


\section{Acknowledgements}
\label{a:ack}
The author would like to express particular gratitude to Michael Kazhdan for advising the research and illuminating discussions. To Apple Inc. for funding my PhD research. To department of Mechanical Engineering at the Johns Hopkins University for departmental fellowship. 

\bibliographystyle{ACM-Reference-Format}
\bibliography{main}

\end{document}